\documentstyle[amssymb,12pt]{article}
\input{tcilatex}
\begin{document}

\begin{flushright}
{\em PPD-IOP-99/16\\}
\end{flushright}
\vskip .1in

\begin{center}
{\Large {\bf On SUSY inspired minimal lepton number violation}}

\vspace{50pt}

{{\bf J. L. Chkareuli }$^{a}$,{\bf \ I. G. Gogoladze }$^{a,b}${\bf , M. G.
Green }$^{c}${\bf , D. E. Hutchcroft }$^{c}${\bf \ and A. B. Kobakhidze } $%
^{a,d}${\\[0pt]
} }

\vspace{6pt}

$^{a}{\em The}$ {\em Andronikashvili Institute of Physics, 380077 Tbilisi,
Georgia \\[0pt]
} $^{b}${\em International Centre for Theoretical Physics, 34100 Trieste,
Italy \\[0pt]
} $^{c}${\em Department of Physics, Royal Holloway, University of London,
Egham, Surrey TW20 0EX, UK }

$^{d}{\em High}$ {\em Energy Physics Division, Department of Physics,
University of Helsinki, 00014 Helsinki, Finland\\[0pt]
}

\vspace{20pt} {\Large {\bf Abstract}}
\end{center}

A minimal lepton number violation (LNV) is proposed which could naturally
appear in SUSY theories, if Yukawa and LNV couplings had a common origin.
According to this idea properly implemented into MSSM with an additional
abelian flavor symmetry the prototype LNV appears due to a mixing of leptons
with superheavy Higgs doublet mediating Yukawa couplings. As a result, all
significant physical manifestations of LNV reduce to those of the effective
trilinear couplings $LL\overline{E}$ and $LQ\overline{D}$ aligned, by size
and orientation in a flavor space, with the down fermion (charged lepton and
down quark ) effective Yukawa couplings, while the effective bilinear terms
appear generically suppressed relative to an ordinary $\mu $-term of MSSM.
Detailed phenomenology of the model related to the flavor-changing processes
both in quark and lepton sectors, radiatively induced neutrino masses and
decays of the LSP is presented. Remarkably, the model can straightforwardly
be extended to a Grand Unified framework and an explicit example with SU(7)
GUT is thoroughly discussed.\thispagestyle{empty}\newpage

\section{Introduction}

Until recently all the confirmed experimental data indicated that lepton and
baryon number are conserved in agreement with Standard Model (SM) of quarks
and leptons. These global conservation laws are in essence an important
by--product of a generic $SU(3)_{C}\otimes SU(2)_{W}\otimes U(1)_{Y}$ gauge
invariance of SM which leads for an ordinary SM particle spectrum together
with a general baryon number conservation to the special conservation laws
for the every known lepton flavor. Therefore, in contrast to the quark case
where the (baryon number conserving) mixing among different generations
occurs, being given by the Cabbibo--Kobayashi--Maskawa (CKM) matrix
elements, leptons cannot mix as an orthodox SM contains the left--handed
massless neutrinos only. However, the recent SuperKamiokande data on
atmospheric neutrinos \cite{2} seem to finally change the situation clearly
indicating that neutrino oscillations ($\nu _{\mu }\rightarrow \nu _{\tau }$
or sterile $\nu _{s}$) actually take place, thus opening a way for new
physics beyond the SM.

On the other hand, strict baryon and lepton number conservation could not
have a firm theoretical foundation unless associated with some local gauge
invariance and, as a result, with new long--distance interactions which are
experimentally excluded~\cite{Lee}. Leaving aside for the moment baryon
number violation (BNV), there could be, in principle, several reasons for
lepton number violation (LNV) related with possible extensions of the SM,
such as a presence of higher--dimensional neutrino mass operators induced by
gravity \cite{grav}, an existence of superheavy right--handed neutrinos
inducing small neutrino masses through the known see--saw mechanism \cite
{see--saw}, new weak triplet Higgs bosons giving tiny masses to neutrinos
directly~\cite{maj} etc. However the minimal supersymmetric extension of the
SM, being in all other aspects also well--motivated, where not only fermions
but also their scalar superpartners automatically become carriers of lepton
(and baryon) numbers, seems to be the most natural and appealing framework
for LNV.

In general, the basic renormalisable dimension--four BNV and LNV couplings
expected in the low--energy MSSM superpotential $W$, unless forbidden by
some side symmetry such as $R$--parity~\cite{Rp}, are given by 
\begin{equation}
\Delta W=\mu _{i}L_{i}H_{u}+\lambda _{ijk}L_{i}L_{j}\overline{E}_{k}+\lambda
_{ijk}^{\prime }L_{i}Q_{j}\overline{D}_{k}+\lambda _{ijk}^{\prime \prime }%
\overline{U}_{i}\overline{D}_{j}\overline{D}_{k}.  \label{1}
\end{equation}
where $i$, $j$, $k$ are generation indices ($i$, $j$, $k=1,2,3$) and an
associated summation is implied (color and weak isospin indices are
suppressed); $L_{i}$($Q_{j}$) denote the lepton (quark) $SU(2)$--doublet
superfields and $\overline{E}_{i}$($\overline{U}_{i}$, $\overline{D}_{i}$)
are $SU(2)$--singlet lepton (up quark, down quark) superfields; $\mu _{i}$
are mass parameters which mix lepton superfields with the ''up'' Higgs
superfield $H_{u}$, while $\lambda _{ijk}$ ($\lambda _{ijk}=-\lambda _{jik}$%
), $\lambda _{ijk}^{\prime }$ and $\lambda _{ijk}^{\prime \prime }$ ($%
\lambda _{ijk}^{\prime \prime }=-\lambda _{ikj}^{\prime \prime }$) are
dimensionless couplings. The first three terms in $\Delta W$ (\ref{1})
violate lepton number, while the last one violates baryon number. Thus there
are 48 new and $a$ $priori${\it \ }unconstrained parameters (beyond those of
the $R$--parity conserving MSSM) with arbitrary flavor structure in general.
Needless to say, this fact presents serious difficulties for a study of the $%
R$--parity violating (RPV) phenomena at a theoretical as well as
phenomenological level (for the recent reviews see \cite{rev} and references
therein).

When estimating possible contributions of RPV interactions to the low energy
processes one typically assumes that only one of the RPV couplings or one
combination of their products is predominant , while the rest are negligibly
small. This assumption made for quarks and leptons taken in the physical
mass basis looks in some cases unnatural with respect to the starting flavor
structure of the couplings involved (\ref{1}) where quarks and leptons
appear in gauge (''unrotated'') basis. Nevertheless, what we have learned
from such studies is that the RPV couplings are typically smaller than the
ordinary gauge couplings, although some of them taken alone could quite be
of order $O(1)$ \cite{rev} for sparticle masses of $O(100)$ $GeV$. Whereas
the lowest individual bounds follow for BNV couplings $\lambda
_{112}^{\prime \prime }\leq 10^{-7}$ and $\lambda _{113}^{\prime \prime
}\leq 10^{-5}$ from double nucleon decay and $n-\overline{n}$ oscillation,
respectively, the strictest combined bounds appear due to the simultaneous
presence of LNV ($\lambda _{ijk}^{\prime }$) and BNV ($\lambda
_{ijk}^{\prime \prime }$) interactions in (\ref{1}) that inevitably leads to
the unacceptably fast proton decay, unless $\lambda _{11k}^{\prime }\lambda
_{11k}^{\prime \prime }\leq 10^{-22}$ (for $k=2,3$) and $\lambda ^{\prime
}\lambda ^{\prime \prime }\leq 10^{-10}$ (for any combination of flavors)
are taken\cite{rev}. So, one could expect, unless somewhat enormous flavor
anisotropy for RPV couplings is assumed that, while SUSY-inspired baryon
number violation might be highly (or even fully) suppressed, lepton number
violation could occur and at a level which seems large enough for possible
observation of many of its spectacular manifestations.

Meanwhile, $R$--parity is not the only symmetry known to ensure proton
stability. In principle, it is not difficult to arrange the general RPV
couplings (\ref{1}) in such a way to have both lepton number violation and
baryon number conservation, i.e. $\mu _{i}\neq 0,$ $\lambda \neq 0,\lambda
^{\prime }\neq 0,$ $\lambda ^{\prime \prime }=0$. For example, it can be
achieved by imposing some discrete symmetry on the quark and lepton fields.
The simplest choice might be the reflection $Z_{2}$ symmetry under which
quark and leptons transform as

\begin{equation}
Q,\text{ }\overline{U},\text{ }\overline{D},\text{ }L,\text{ }\overline{E}%
\rightarrow -Q,-\overline{U},-\overline{D},\text{ }L,\text{ }\overline{E}%
\text{ .}  \label{1a}
\end{equation}
However, the $Z_{3}$ symmetry

\begin{equation}
Q,\text{ }\overline{U},\text{ }\overline{D},\text{ }L,\text{ }\overline{E}%
\rightarrow Q,\text{ }\omega ^{2}\overline{U},\text{ }\omega \overline{D},%
\text{ }\omega ^{2}L,\text{ }\omega ^{2}\overline{E}\text{ \quad ( }\omega
=e^{i2\pi /3}\text{ ) ,}  \label{1b}
\end{equation}
known as the baryon parity\cite{Bpar}, seems to be of particular interest.
The reason is that the baryon parity happens to be the
superstring--inherited gauge discrete symmetry~which is stable under
gravitational corrections. This symmetry, as was shown\cite{Bpar}, forbids
not only the dimension--four BNV interactions in (\ref{1}), but still
dangerous dimension--five operators as well. At the same time, acting more
selectively than $R$--parity, $Z_{3}$ baryon parity allows all LNV
interactions in (\ref{1}). In this connection it seems reasonable to
suppress fully all BNV couplings ($\lambda ^{\prime \prime }=0$) and be
focused further on LNV only. From here on we assume that it is the case when
considering MSSM.

However, the situation becomes further complicated when one tries to embed
MSSM with LNV couplings (\ref{1}) into the more fundamental framework of a
Grand Unified Theory (GUT) which typically keeps quarks and leptons in
common multiplets. By contrast, discrete symmetries protecting the proton
stability, such as $Z_{3}$ baryon parity mentioned, treat quarks and leptons
differently and hence they come into conflict with the known minimal GUTs.
Nevertheless, a number of the properly extended GUTs have been constructed 
\cite{GUTS} where the low-energy lepton number violation versus baryon
number conservation appears once the starting symmetry breaks down to MSSM.

An excessively wide variety of the possible LNV couplings (36 of the $%
\lambda $ and $\lambda ^{\prime }$ terms) brings up another critical point:
what is a basic prototype LNV form which could naturally appear in MSSM? The
bilinear terms $\mu _{i}L_{i}H_{u}$ in the superpotential $\Delta W$ (\ref{1}%
) might be such a minimal possibility, if all trilinear couplings were
generically absent (i.e. $\lambda =\lambda ^{\prime }=0$). These trilinear
couplings could be prohibited by some additional symmetry -- typically by a
gauge (ordinary or anomalous) $U(1)$ symmetry concerning the quark and
lepton flavors \cite{Dudas} in the properly extended MSSM. The bilinear
terms in themselves can be rotated away, thus recovering the effective
trilinear $\lambda $ and $\lambda ^{\prime }$ terms from the ordinary Yukawa
couplings for leptons and down quarks, respectively. Such a type of minimal
theory with a generic alignment between LNV terms and Yukawa couplings was
considered in a number of papers [9-11] and many of its interesting features
had been established. The most appealing one is that the size and flavor
structure of LNV couplings are essentially given now by those of Yukawa
couplings so as to naturally overcome all the currently available
experimental constraints (among them the most stringent ones which might
follow from $K^{0}-\overline{K}^{0}$ mixing, $K^{0}\rightarrow e_{i}%
\overline{e}_{j}$ decays etc.).

However, this model has, at least, two serious drawbacks. The first one
concerns the natural size of new, arbitrary in general, mass parameters $\mu
_{i}$ relative to the basic MSSM mass given by an ordinary $\mu $-term , $%
\mu H_{u}\overline{H}_{d}$. All the $\mu _{i}$ certainly must be arranged to
be at most of order of $\mu $ not to disturb significantly an electroweak
symmetry breaking in MSSM. Furthermore, when extending to the GUT framework
the bilinear terms lead in general, together with the lepton mixing with a
weak Higgs doublet, to the quark mixing with a color Higgs triplet, thus
inducing baryon number violation as well. Again, the only possible solution
to the problem might use the electroweak scale order masses $\mu _{i}$ in
the GUT symmetry-invariant bilinear couplings. This would require that new
fine-tuning conditions, besides a notorious gauge hierarchy one, should be
satisfied in a very ad hoc way.

The second problem is that when the SUSY soft breaking terms are included
the bilinear couplings cannot be rotated away and lead to an enormously
complicated scalar sector with sneutrinos condensed, and neutrinos and
neutralinos mixed by a proper (seven--by--seven in a general $\widetilde{%
\gamma }-\widetilde{Z}-\widetilde{H}_{u}^{0}-\widetilde{H}_{d}^{0}-\nu
_{e}-\nu _{\mu }-\nu _{\tau }$ basis) mass matrix. As a result, there appear
the tree-level neutrino masses which are generally too large unless some
precise alignment between the bilinear LNV and corresponding SUSY
soft--breaking terms is provided. There was made some progress in recent
years towards his problem in the framework of supergravity theories \cite
{add1}.

Nevertheless, the Higgs-lepton mixing model (or HLM model hereafter) of LNV
seems to be, against all the odds, the most appealing one as it uniquely
links with flavour physics peculiarities both in quark and lepton sectors.
We take it as the starting point towards a new framework related with a
possible common origin of the LNV and Yukawa couplings. While our model is
also based on a generic Higgs-lepton mixing, we propose, in contrast to an
ordinary picture (\ref{1}), that this mixing appears not with a standard
MSSM Higgs doublet $H_{u}$ but with some superheavy weak doublets $\Phi +%
\overline{\Phi }$ mediating the down fermion (down quark and charged lepton)
effective Yukawa couplings. This mechanism is readily extended to the GUT\
framework as well, since it allows any large $\Phi -L$ mixing masses. As a
result, we drive at another picture where only effective trilinear LNV
couplings appear being aligned with the Yukawa ones, whereas the ordinary
bilinear $H_{u}-L$ mixings (\ref{1}) are proved to be strongly suppressed.
So, the model considered can be qualified as a purely trilinear model or,
equally (if expressed in terms of primary couplings, see below), as the
heavy Higgs-lepton mixing (HHLM) model of LNV.

Depending on the $U(1)_{A}$ charges assigned to matter and Higgs superfields
involved one can come to an ordinary HLM model or the HHLM model considered
here. Some of the predictions of both models, particularly those concerning
quark flavor conservation in the LNV inspired processes are very similar.
However, there are principal differences as well. The point is that an
influence of the SUSY soft breaking sector, being predominant for an
ordinary HLM, is quite negligible for the HHLM model. Therefore, the
LNV-Yukawa alignment, while appeared in both of models, leads just in the
latter case to the distinctive relations between various LNV processes
arising from the slepton and squark exchanges, which are basically
conditioned by quark and lepton mass hierarchy (see Section III). By
contrast, in an ordinary HLM case these processes appear to be essentially
determined by $W$ and $\ Z$ bozon exchanges and, as a result, are largely
flavor-independent. Whilst at the moment one can not phenomenologically
distinguish them, further extension to the GUT framework seems, as we argue
below, to favor the HHLM model.

We construct an explicit example of the R-parity violating $SU(7)$ GUT\cite
{FFF}. The preference given to $SU(7)$ model over other grand unified
frameworks is essentially determined by the missing VEV solution to the
doublet-triplet splitting problem which naturally appears in $SU(N)$ GUTs
starting from the $SU(7)$\cite{JGK}. In this model the effective LNV
couplings with the Yukawa-aligned structure immediately evolve, while the
baryon number non-conserving RPV couplings are safely projected out by the
proper missing VEV vacuum configuration that breaks the starting $SU(7)$
symmetry down to the one of MSSM. This implies that a missing VEV solution
to gauge hierarchy problem can generically protect baryon number
conservation in the RPV $SU(7)$ SUSY GUT.

The rest of the paper is organized as follows. In Section II we present the
proposed HHLM model of minimal LNV in MSSM whose a detailed phenomenological
study will be given in Section III. In Section IV the $SU(7)$ framework for
lepton number violation versus baryon number conservation is presented.
Finally, in Section V our conclusions are summarized.

\section{Minimal lepton number violation in MSSM}

We argue here that the proposed HHLM model with a heavy Higgs-lepton mixing
is in fact the minimal generic form of LNV whose basic predictions are
related with masses and mixings of quarks and leptons.

Following to the observed up-down hierarchy in quark mass spectrum, where
the top mass is clearly leading, we propose that, while the up quark Yukawa
couplings have a usual trilinear form 
\begin{equation}
W_{U}=Y_{jk}^{u}Q_{j}\overline{U}_{k}H_{u}  \label{a}
\end{equation}
the down fermion (down quark and charged lepton) Yukawa couplings could have
one more dimension being generically mediated by superheavy Higgs doublets $%
\Phi +\overline{\Phi }$ which simultaneously mix with a standard ''down''
Higgs\ $\overline{H}_{d}$ and leptons $L_{i}$:

\begin{eqnarray}
W_{D} &=&G_{jk}^{d}Q_{j}\overline{D}_{k}\overline{\Phi }+G_{jk}^{l}L_{j}%
\overline{E}_{k}\overline{\Phi }+f\Phi \overline{H}_{d}S+M_{\Phi }\overline{%
\Phi }\Phi \text{ ,}  \label{b} \\
W_{LNV} &=&h_{i}L_{i}\Phi T  \label{c}
\end{eqnarray}
($Y_{jk}^{u}$, $G_{jk}^{l,d}$, $f$ and $h_{i}$ are dimensionless coupling
constants properly introduced). The couplings (\ref{a}-\ref{c}) are in
substance the most general ones which can appear in MSSM complemented by
some abelian flavor symmetry as an anomalous $U(1)_{A}$ symmetry in the case
considered which is supposed to properly arrange the rest of hierarchy in
quark and lepton mass spectra \cite{col}. We have introduced the singlet
scalar superfields $S$ and $T$ , the basic carriers of the $U(1)_{A}$
charge. They are presumed to develop the high scale (up to string scale
order) VEVs through the Fayet-Iliopoulos $D$-term related with $U(1)_{A}$
symmetry \cite{witten}. In terms of Q$_{A}^{S}$ and Q$_{A}^{T}$ all the
other charges appear to be properly expressed so that the direct
Higgs-lepton mixing terms $\mu _{i}L_{i}H_{u}$ (\ref{1}) are strictly
prohibited until $U(1)_{A}$ symmetry breaks\footnote{%
We introduced two singlet scalar superfields $S$ and $T$ in order to have an
ordinary $\mu $--term $\mu H_{u}\overline{H}_{d}$ in superpotential, while
the bilinear LNV terms $\mu _{i}L_{i}H_{u}$ are still prohibited (a minimal
case with one scalar superfield $S$ would admit both of terms in
superpotential). In $SU(7)$ GUT framework considered in Section IV they
appear as the non-trivial $SU(7)$ multiplets.}. Such a mixing is allowed
only with superheavy Higgs doublet system $\Phi +\overline{\Phi }$ having
Planck scale order mass, $M_{\Phi }=O(M_{P})$.

Once $U(1)_{A}$ symmetry breaks ($<S>$ $\neq 0$, $<T>$ $\neq 0$) we come to
the common mass matrix of all Higgs and lepton superfields involved:

\bigskip

\begin{equation}
\begin{tabular}{ccccccccccc}
${\bf \looparrowright }$ &  &  &  & ${\bf L}_{i}$ &  &  & $\overline{{\bf H}}%
_{d}$ &  &  & $\overline{{\bf \Phi }}$ \\ 
${\bf L}_{i}$ &  &  &  & $0$ &  &  & $0$ &  &  & $0$ \\ 
${\bf H}_{u}$ &  &  &  & $0$ &  &  & $\mu $ &  &  & $0$ \\ 
${\bf \Phi }$ &  &  &  & $h_{i}<T>$ &  &  & $f<S>$ &  &  & $M_{\Phi \text{ }%
} $%
\end{tabular}
\label{d}
\end{equation}
whose diagonalization up to the second order mixing terms leads to the
effective down fermion Yukawa and LNV couplings of type (keeping the same
notation for the ''rotated'' superfields).

\begin{eqnarray}
W_{D}^{eff} &=&G_{jk}^{d}f\frac{<S>}{M_{\Phi }}Q_{j}\overline{D}_{k}%
\overline{H}_{d}+G_{jk}^{l}f\frac{<S>}{M_{\Phi }}L_{j}\overline{E}_{k}%
\overline{H}_{d}  \label{e} \\
W_{LNV}^{eff} &=&h_{i}\frac{<T>}{M_{\Phi }}[G_{jk}^{l}L_{i}L_{j}\overline{E}%
_{k}+G_{jk}^{d}L_{i}Q_{j}\overline{D}_{k}+\frac{<S>}{M_{\Phi }}\mu L_{i}H_{u}%
]  \label{f}
\end{eqnarray}
Thereby, the effective coupling constants included in Eqs. (\ref{e}, \ref{f}%
) are read off as

\begin{equation}
Y_{jk}^{l}=fG_{jk}^{l}\frac{<S>}{M_{\Phi }}\text{, \quad }%
Y_{jk}^{d}=fG_{jk}^{d}\frac{<S>}{M_{\Phi }}  \label{j4}
\end{equation}
and 
\begin{equation}
\lambda _{ijk}=h_{i}G_{jk}^{l}\frac{<T>}{M_{\Phi }}\text{, \quad }\lambda
_{ijk}^{\prime }=h_{i}G_{jk}^{d}\frac{<T>}{M_{\Phi }}  \label{j5}
\end{equation}
from which the basic LNV-Yukawa coupling relations immediately follows 
\begin{equation}
\lambda _{ijk}=\epsilon _{i}Y_{jk}^{l}\text{ , \quad }\lambda _{ijk}^{\prime
}=\epsilon _{i}Y_{jk}^{d}\text{\ }  \label{2}
\end{equation}
with the proportionality parameters $\epsilon _{i}$ determined as

\begin{equation}
\epsilon _{i}=\frac{h_{i}}{f}\frac{<T>}{<S>}\text{ .}  \label{ss}
\end{equation}

Remarkably, as one can see from Eqs. (\ref{f}) and (\ref{ss}), the effective 
$L-H_{u}$ mixing masses $\mu _{i}$ appear generically related to the basic
MSSM\ mass $\mu $

\begin{equation}
\mu _{i}=f\epsilon _{i}(\frac{<S>}{M_{\Phi }})^{2}\mu \text{ ,}  \label{sss}
\end{equation}
while being properly suppressed. When taking $\frac{<T>}{M_{\Phi }}\sim 
\frac{<S>}{M_{\Phi }}\sim 10^{-2}$ to provide the observed up-down mass
hierarchy (or, equally, the observed mass scale for the $b$ quark and $\tau $
lepton) in Yukawa coupling constants (\ref{j4}) one comes to a natural bound
for $\mu _{i}$ parameters, $\mu _{i}\lesssim 0.01$ $GeV$ for $\mu =O(100$ $%
GeV)$. This is in fact too small to have any sizeable influence on the
scalar sector and tree--level neutrino masses, as in a bilinear HLM case
mentioned above. Therefore, as to the significant physical manifestations of
LNV, one has only those related with the effective trilinear couplings (\ref
{f}) being aligned (by size and orientation in flavor space) with down
fermion Yukawa couplings (\ref{f})\footnote{%
There can appear, for the effective Yukawa and LNV operators mediated by the
superheavy Higgs doublet pair $\Phi +\overline{\Phi }$, the competitive
gravitational corrections smearing out the alignment conditions (\ref{2}).
However, they will be absent if the bilinear $\Phi \overline{\Phi }$ has a
non-zero $U(1)_{A}$ charge, thus getting mass once the $U(1)_{A}$ symmetry
breaks. Indeed, if so, all the high-dimension operators which appear through
the $\Phi +\overline{\Phi }$ exchanges are not neutral under $U(1)_{A}$ and,
therefore, can not be induced by gravity.}.

And the last is that the model also allows a straightforward extension to a
GUT, particularly, to the $SU(7)$ GUT with a natural solution to the
hierarchy problem due to the basic vacuum configuration appeared with no
VEVs along all the color directions. There the singlet scalar superfields $S$
and $T$ are replaced, respectively, by one of the fundamental Higgs
multiplets which break $SU(7)$ to $SU(5)$ and a basic adjoint multiplet of $%
SU(7)$ which just projects out all the BNV couplings and leaves the LNV ones
only. Thus, the extra scalar superfields specially introduced in MSSM
framework are turned out to naturally exist in the $SU(7)$ GUT. We consider
this in more detail in Section IV.

\section{Phenomenology of HHLM model}

During the last few years the SUSY inspired baryon and lepton number
non-conservation has called a considerable attention. As a result, an
extensive study of the bounds from various low--energy processes on RPV
couplings (\ref{1}) was carried out, as well as many specific manifestations
of RPV interactions at present and future colliders were investigated (see
[7, 17] and references therein). In this Section we consider some of the
immediate consequences of lepton number violation which could emerge in the
minimal HHLM (Heavy Higgs-Lepton Mixing) model proposed. We pursue this
study on the purely phenomenological level as given by the basic
Yukawa-aligned trilinear LNV interactions (see Eq. (\ref{2})) with no
significant bilinear terms generated in the soft SUSY breaking sector either
on tree level (see Eq. (\ref{sss})) or through the radiative corrections.

\subsection{LNV couplings in physical basis}

Towards this end one must go in the general alignment conditions (\ref{2})
to the physical (mass) basis where the down fermion Yukawa matrices $Y^{l}$
and $Y^{d}$ are diagonal\footnote{%
We have assumed that the fermion and sfermion mass matrices can
simultaneously be brought to the diagonal form as is usually taken in
ordinary MSSM.}. As a result, the basic relations between the effective $%
\lambda $ and $\lambda ^{\prime }$ couplings and masses of down fermions
follow, which are

\begin{equation}
\lambda _{ijk}=\frac{\epsilon _{i}}{Vc_{\beta }}\left( {\ 
\begin{array}{ccc}
\,\,m_{e}~~ & \,~~ & \,\, \\ 
~~ & m_{\mu }\,\,~~ & \,\, \\ 
\,\,~~ & \,\,~~ & \,\,m_{\tau }
\end{array}
}\right) _{jk}  \label{u}
\end{equation}
and

\begin{equation}
\lambda _{ijk}^{^{\prime }}=\frac{\epsilon _{i}}{Vc_{\beta }}{\left( 
\begin{array}{ccc}
\,\,m_{d}~~ & \,~~ & \,\, \\ 
~~ & m_{s}\,\,~~ & \,\, \\ 
\,\,~~ & \,\,~~ & \,\,m_{b}
\end{array}
\right) }_{jk}\text{,}  \label{w}
\end{equation}
respectively (here $V\approx 174$ $GeV$ is the electroweak VEV, while $\tan
\beta $ is an usual ratio of the VEVs of ''up'' and ''down'' Higgses $H_{u}$
and $\overline{H}_{d}$). There appear clear in physical basis some of the
generic features of the HHLM model which essentially determine its
phenomenological implications considered further in this Section.

The first and foremost is that, while both of basic LNV couplings $L_{i}L_{j}%
\overline{E}_{k}$ and $L_{i}Q_{j}\overline{D}_{k}$ generally violate lepton
number $\left| \Delta N^{L}\right| $ $=1$ and conserve baryon (quark) number 
$\Delta N^{Q}=0$, some additional selection rules related with flavor
(generation) species of leptons and quarks come into play in the HHLM model:

\begin{itemize}
\item  A partial lepton flavor conservation according which only one of
lepton flavor numbers is violated at a time, while the other two are still
retained 
\begin{equation}
\left| \Delta N^{L_{i}}\right| =\delta _{im}\text{ \quad (}i,m=1,2,3\text{)
, }  \label{bu}
\end{equation}

\item  An exact quark flavor conservation 
\begin{equation}
\Delta N^{Q_{i}}=0\text{ .}  \label{ob}
\end{equation}
\end{itemize}

Next is a large reduction of the possible LNV couplings being conditioned by
the selection rules (\ref{bu}) and (\ref{ob}). One can quickly confirm that
only $\lambda $ and $\lambda ^{\prime }$ couplings with the last two indices
equal are left in the physical basis. This gives $6+9=15$ physical $\lambda $
and $\lambda ^{\prime }$ couplings in total (instead of $9+27=36$ as in a
general case) depending, in effect, on three unknown parameters $\epsilon
_{i}$ only.

And the last (but certainly not least as it will be seen below) is a natural
smallness of LNV coupling constants as they are seen from Eqs. (\ref{u}, \ref
{w}) being largely determined by the known masses of leptons and quarks

\begin{eqnarray}
\lambda _{211} &=&\frac{\epsilon _{3}}{\epsilon _{2}}\lambda _{311}\simeq
2.9\cdot 10^{-6}\frac{\epsilon _{2}}{c_{\beta }}  \nonumber \\
\lambda _{122} &=&\frac{\epsilon _{3}}{\epsilon _{1}}\lambda _{322}\simeq
6.4\cdot 10^{-4}\frac{\epsilon _{1}}{c_{\beta }}  \nonumber \\
\lambda _{133} &=&\frac{\epsilon _{2}}{\epsilon _{1}}\lambda _{233}\simeq
1.1\cdot 10^{-2}\frac{\epsilon _{1}}{c_{\beta }}  \label{11} \\
&&  \nonumber \\
&&  \nonumber \\
\lambda _{i11}^{\prime } &=&(2.9\div 8.6)\cdot 10^{-5}\frac{\epsilon _{i}}{%
c_{\beta }}  \nonumber \\
\lambda _{i22}^{\prime } &=&(0.6\div 1.7)\cdot 10^{-3}\frac{\epsilon _{i}}{%
c_{\beta }}  \nonumber \\
\lambda _{i33}^{\prime } &=&(2.4\div 2.6)\cdot 10^{-2}\frac{\epsilon _{i}}{%
c_{\beta }}  \label{12}
\end{eqnarray}
where the numerical values shown follow from the masses of leptons and
quarks (including the corresponding uncertainties in the down quark masses $%
m_{d}=5\div 15$ $MeV$, $m_{s}=100\div 300$ $MeV$ and $m_{b}=4.1\div 4.5$ $GeV
$ \cite{PDG}). The unknown parameters $\epsilon _{i}$ in (\ref{11}) and (\ref
{12}) can be all of the same order or follow to some hierarchy dictated by
flavor symmetry of the underlying theory, if it is the case [10, 19].

\subsection{Constraints from low--energy processes}

\noindent Consider first the possible effective LNV interactions which
follow from still unconstrained primary couplings (\ref{1}) when all the
intermediate sleptons and squarks are integrated out. The typical
four--fermion operators appeared are listed in TABLE I. Generally, these
operators mediating the possible flavor-changing transitions both in quark
and lepton sectors could contribute to known processes leading to the
deviations from the observed rates of $K^{0}-\overline{K}^{0}$ and $B^{0}-%
\overline{B}^{0}$ oscillations, charged current universality, $e-\mu -\tau $
universality, atomic parity violation and others. Also they could induce the
rare decays of mesons and leptons many of which are highly suppressed or
even forbidden in the SM. Using all the related data and observations
presently existed one extracts rather severe bounds on the LNV couplings $%
\lambda $ and $\lambda ^{\prime }$ and/or on their products [7, 17].

Now, going from general case to the HHLM model with a generic LNV-Yukawa
alignment presented in physical basis by the selection rules (\ref{bu}, \ref
{ob}) one can immediately confirm that there are no significant LNV
contributions to any of the flavor-changing neutral current (FCNC)
processes. The usually dangerous tree level LNV processes, such as leptonic
and semi-leptonic decays of pseudoscalar mesons ($K^{0}\rightarrow e_{i}%
\overline{e}_{j}$, $K^{+}\rightarrow \pi ^{+}e_{i}\overline{e}_{j}$, $%
K^{+}\rightarrow \pi ^{+}\nu _{i}\overline{\nu }_{j}$, $B^{0}\rightarrow
e_{i}\overline{e}_{j}$, $B\rightarrow X_{q}\nu _{i}\overline{\nu }_{j}$) as
well as the possible LNV contributions to $K^{0}-\overline{K}^{0}$ and $%
B^{0}-\overline{B}^{0}$ oscillations, are naturally forbidden in our
scenario. Moreover, since the typical strength of the LNV couplings in the
HHLM model (see Eqs. (\ref{11}, \ref{12})) is smaller than a strength of
electroweak interactions, even the loop contributions to the above processes
(to those with $i=j$) are largely dominated by the usual SM interactions.

At the same time in cases when the LNV induced processes are allowed in HHLM
model they appear to readily satisfy the existing bounds due a generic
smallness of the LNV couplings appeared, thus leading to the quite
acceptable limitations on the $\epsilon $-parameters and $\tan \beta $
involved (\ref{11}, \ref{12}). For example, one of the most stringent bounds
that can be extracted from the atomic transition conversion process\cite{rev}
$\mu ^{-}+$ $^{45}Ti\rightarrow e^{-}+$~$^{45}Ti$ gives a bound $\frac{%
\epsilon _{1}\epsilon _{2}}{c_{\beta }^{2}}<1$ for the mediating sfermion
masses of $m_{\widetilde{f}}=300$ $GeV$.

Another important constraint comes from the current experimental limits on
neutrinoless double beta decay\cite{rev}. This process is triggered by the
six--fermion effective operators of the form $eu\overline{d}eu\overline{d}$
which appear in the low--energy theory after one integrates out sleptons and
squarks. The analysis of the disintegration process of $^{76}Ge$ (with
half--life time $T_{\frac{1}{2}}>1.1\cdot 10^{25}$ years) leads to the bound 
$\frac{\epsilon _{1}}{c_{\beta }}\lesssim 1.1$.

For the $\epsilon $-parameters fixed, one can further get the bound on $\tan
\beta $ preferably according to the largest LNV couplings $\lambda _{i33}$
and $\lambda _{i33}^{\prime }$ appeared (\ref{11}, \ref{12}) . The most
stringent bound comes from the charged current universality constraints in $%
\tau $--decays which imply $\tan \beta \lesssim 6$, if $\epsilon _{2,3}\sim
1 $ is taken.

\subsection{Three-body leptons decays}

\noindent The rare decays of leptons, along with the processes mentioned
above, are usually treated as the most promising ones in searching for
lepton flavor violation phenomenon \cite{exp}. Here we consider three--body
leptons decays of $\mu $ and $\tau $ triggered by the last of four--fermion
operators listed in TABLE I. All these processes can be presented by a
generic transition of type $e_{j}\rightarrow e_{k}+e_{l}+\overline{e}_{m}$
which proceeds by the exchange of a sneutrino $\widetilde{\nu }_{i}$ in the $%
t$ as well as $u$ channel. The effective Lagrangian can be expressed as \cite
{declep} 
\begin{eqnarray}
{\cal L}(e_{j} &\rightarrow &e_{k}+e_{l}+\overline{e}_{m})={\cal F}_{jklm}%
\overline{e}_{kR}e_{jL}\overline{e}_{lL}e_{mR}+{\cal F}_{mlkj}\overline{e}%
_{kL}e_{jR}\overline{e}_{lR}e_{mL}  \nonumber \\
&&+{\cal F}_{jlkm}\overline{e}_{lR}e_{jL}\overline{e}_{kL}e_{mR}+{\cal F}%
_{mklj}\overline{e}_{lL}e_{jR}\overline{e}_{kR}e_{mL}  \label{13}
\end{eqnarray}
where 
\begin{equation}
{\cal F}_{jklm}=\sum_{i}\left( \frac{1}{m_{\widetilde{\nu }_{i}}^{2}}\right)
\lambda _{ijk}\lambda _{ilm}^{\star }.  \label{14}
\end{equation}
The first two terms in (\ref{13}) correspond to $t$ channel exchange
diagrams, while the last two terms correspond to $u$ channel exchange
diagrams.

Substituting the couplings from (\ref{11}) into (\ref{14}), we have
calculated the branching ratios of various decays of $\mu $ and $\tau $.
They are exposed in TABLE II. Unless considerable experimental progress is
made the most of branchings from TABLE II are clearly too small for the
corresponding decays to be detected in a near future. The only exclusion
might exist for the dominant $\tau $ decay mode $\tau \rightarrow 3\mu $,
although even for the favorable parameter area with $\epsilon _{2,3}\sim 1$, 
$\tan \beta =6$ and sneutrino masses of 100 $GeV$, its branching is turned
out to be $Br(\tau \rightarrow 3\mu )\approx 10^{-8}$. Meanwhile, although
presently one hardly expects a sensitivity better than $10^{-7}$ for any
rare decay mode of $\tau $ \cite{exp}, there is planned at LHC to reach the
branching level $10^{-9}$\cite{exp*}, particularly, for the mode mentioned.

In this connection the model predicts some of the quite specific signals
that could be tested at LHC or other future facilities such as a muon
collider and $\tau $--factories. The first is that according to the flavor
selection rule (\ref{ob}) the 3-lepton $\tau $ decays with the identical
di-leptons in final state, like as $\tau ^{\mp }\rightarrow e^{\mp }e^{\mp
}\mu ^{\pm }$ and $\tau ^{\mp }\rightarrow \mu ^{\mp }\mu ^{\mp }e^{\pm }$,
are strictly prohibited. The second is the characteristic relations appeared
among the partial decay widths which can be expressed through the lepton
masses and $\epsilon _{i}$ parameters in a following way 
\begin{eqnarray}
\frac{\Gamma (\mu \rightarrow 3e)}{\Gamma (\tau \rightarrow 3e)} &\simeq
&\left( \frac{\epsilon _{2}m_{\mu }}{\epsilon _{3}m_{\tau }}\right)
^{2}\cdot \rho  \nonumber \\
\frac{\Gamma (\tau \rightarrow 3e)}{\Gamma (\tau \rightarrow 3\mu )}
&=&\left( \frac{\epsilon _{1}m_{e}}{\epsilon _{2}m_{\mu }}\right) ^{2}
\label{15} \\
\frac{\Gamma (\tau \rightarrow 3e)}{\Gamma (\tau \rightarrow e\mu \overline{%
\mu })} &=&\left( \frac{m_{e}}{m_{\mu }}\right) ^{2}  \nonumber \\
\frac{\Gamma (\tau \rightarrow \mu e\overline{e})}{\Gamma (\tau \rightarrow
3\mu )} &=&\left( \frac{m_{e}}{m_{\mu }}\right) ^{2}  \nonumber
\end{eqnarray}
together with an elegant ''model-independent'' branching combination 
\begin{equation}
\frac{\Gamma (\tau \rightarrow 3e)\Gamma (\tau \rightarrow 3\mu )}{\Gamma
(\tau \rightarrow e\mu \overline{\mu })\Gamma (\tau \rightarrow \mu e%
\overline{e})}=1  \label{16}
\end{equation}
containing neither lepton masses nor the $\epsilon _{i}$ parameters (in the
first relation in (\ref{15}) the approximately equal masses for the second
and third generation sneutrinos were taken for simplicity; $\rho $ stands
for the phase volume factor, $\rho \simeq (m_{\mu }/m_{\tau })^{5}$).

One can readily confirm that as the distinctive suppression of some of $\tau 
$ decay modes mentioned, so the above strict relations between its modes
allowed appear as a direct consequence of the Yukawa-aligned structure of
the LNV $\lambda $ couplings clearly manifested itself when taking in a
physical basis (\ref{u}).

\subsection{LSP decays}

The main predictions of the HHLM model proposed certainly belong to the
lightest neutralino (LSP) decays\footnote{%
We consider the lightest neutralino as the LSP.}, which could, in principle,
be tested even at the currently working facilities if the sparticle masses
were in a proper area.

Recall that the LNV decays of the LSP drastically changes a standard
missing--energy signature being in the ordinary $R$--parity conserving MSSM.
Instead, the LSP gives rise to the high-energy particles all of which but
neutrinos can be detected directly. These decays are in effect the two-step
processes being properly mediated by sleptons and/or squarks. At the first
step the LSP freely goes to lepton-slepton (quark-squark) pair and then
slepton (squark) decays through the proper LNV coupling into final lepton
(or quark-lepton) pair. While the first step is practically
flavor-independent (since the LSP goes to all fermion-sfermion pair but
possibly the top-stop system mainly due to its electroweak eigenstate
components photino $\widetilde{\gamma }$ and zino $\widetilde{Z}$), the
second one, according to the basic coupling equations (\ref{u}) and (\ref{w}%
), crucially depends on the lepton and /or quark species.

As a result, only some particular (distinctively configured in a flavor
space by the selection rules (\ref{bu}) and (\ref{ob})) LSP decay modes
appear. For the leptonic and semi-leptonic LSP decays proceeding in the
final LNV stage through the $\lambda $ and $\lambda ^{\prime }$ couplings,
respectively, they are

\begin{eqnarray}
\chi _{1}^{0} &\rightarrow &e_{i}+\overline{e}_{k}+\nu _{k}\text{ ,}
\label{l1} \\
\chi _{1}^{0} &\rightarrow &\nu _{i}+\overline{e}_{k}+e_{k}  \label{l2}
\end{eqnarray}
and

\begin{eqnarray}
\chi _{1}^{0} &\rightarrow &e_{i}+\overline{d}_{k}+u_{k}\text{ ,}  \label{l3}
\\
\chi _{1}^{0} &\rightarrow &\nu _{i}+\overline{d}_{k}+d_{k}  \label{l4}
\end{eqnarray}
where indices $i$ and $k$ shown correspond to the lepton and quark
generation species appeared. In this connection the most experimentally
interesting cases are the decays (\ref{l1}) and (\ref{l3}) whose branchings
are essentially determined by flavors of the charged leptons (and down
quarks) involved: 
\begin{eqnarray}
\frac{\Gamma (\chi _{1}^{0}\rightarrow e_{i}\overline{e}_{k}\nu _{k})}{%
\Gamma (\chi _{1}^{0}\rightarrow e_{j}\overline{e}_{m}\nu _{m})} &\simeq
&\left( \frac{\epsilon _{i}m_{e_{k}}}{\epsilon _{j}m_{e_{m}}}\right) ^{2}
\label{lsprel} \\
\frac{\Gamma (\chi _{1}^{0}\rightarrow e_{i}\overline{d}_{k}u_{k})}{\Gamma
(\chi _{1}^{0}\rightarrow e_{j}\overline{d}_{m}u_{m})} &\simeq &\left( \frac{%
\epsilon _{i}m_{d_{k}}}{\epsilon _{j}m_{d_{m}}}\right) ^{2}  \nonumber
\end{eqnarray}
We have assumed for simplicity, while deriving these relations, that all
generation sleptons and squarks mediating decay processes (\ref{l1}-\ref{l4}%
) have approximately the same masses; also the CKM mixing was neglected for
quarks.

>From these general relations a variety of particular ones follow when taking
some special orientation of flavor indices. Among them are cases when $i=j$ (%
$k\neq m$) and $k=m$($i\neq j$), respectively,

\begin{eqnarray}
\frac{\Gamma (\chi _{1}^{0}\rightarrow e_{i}\overline{e}_{k}\nu _{k})}{%
\Gamma (\chi _{1}^{0}\rightarrow e_{i}\overline{e}_{m}\nu _{m})} &\simeq
&\left( \frac{m_{e_{k}}}{m_{e_{m}}}\right) ^{2}\text{ ,}  \label{q} \\
\frac{\Gamma (\chi _{1}^{0}\rightarrow e_{i}\overline{e}_{k}\nu _{k})}{%
\Gamma (\chi _{1}^{0}\rightarrow e_{j}\overline{e}_{k}\nu _{k})} &\simeq
&\left( \frac{\epsilon _{i}}{\epsilon _{j}}\right) ^{2}  \label{p}
\end{eqnarray}
for leptonic decay modes and

\begin{eqnarray}
\frac{\Gamma (\chi _{1}^{0}\rightarrow e_{i}\overline{d}_{k}u_{k})}{\Gamma
(\chi _{1}^{0}\rightarrow e_{i}\overline{d}_{m}u_{m})} &\simeq &\left( \frac{%
m_{d_{k}}}{m_{d_{m}}}\right) ^{2}\text{ ,}  \label{qq} \\
\frac{\Gamma (\chi _{1}^{0}\rightarrow e_{i}\overline{d}_{k}u_{k})}{\Gamma
(\chi _{1}^{0}\rightarrow e_{j}\overline{d}_{k}u_{k})} &\simeq &\left( \frac{%
\epsilon _{i}}{\epsilon _{j}}\right) ^{2}  \label{pp}
\end{eqnarray}
for the semi-leptonic ones, as well as one more relation connecting both of
sets

\begin{equation}
\frac{\Gamma (\chi _{1}^{0}\rightarrow e_{i}\overline{e}_{k}\nu _{k})}{%
\Gamma (\chi _{1}^{0}\rightarrow e_{j}\overline{e}_{k}\nu _{k})}\simeq \frac{%
\Gamma (\chi _{1}^{0}\rightarrow e_{i}\overline{d}_{k}u_{k})}{\Gamma (\chi
_{1}^{0}\rightarrow e_{j}\overline{d}_{k}u_{k})}  \label{z}
\end{equation}
which just like as the $\tau $ lepton branching relation (\ref{16}) contains
no any parameter at all.

The basic signature of the LSP decays, as it can quickly be read off the
relations (\ref{lsprel}), (\ref{q}) and (\ref{qq}), is an overdominance of
the modes with the heaviest lepton and quark families, clearly manifested
when comparing the cases with the same charged lepton taken. Among them the
semileptonic modes $\chi _{1}^{0}\rightarrow e_{i}\overline{b}t$ are, of
course, dominant, if LSP heavier than top quark. Otherwise, the leptonic
modes $\chi _{1}^{0}\rightarrow e_{i}\overline{\tau }\nu _{\tau }$ dominate.
Should the class of decays (\ref{l2}) and \ref{l4}), no tagged by charged
leptons, is considered the modes $\chi _{1}^{0}\rightarrow \nu _{i}\overline{%
b}b$ and $\chi _{1}^{0}\rightarrow \nu _{i}\overline{\tau }\tau $ appear to
be leading. On the other hand, which class of decays from the above two
dominates strongly depends on the nature of the LSP by itself \cite{mor}. If
the LSP mainly consist of the $\widetilde{\gamma }$ component, branching
fractions to the modes (\ref{l1}, \ref{l3}) are large, while the dominant $%
\widetilde{Z}$ component gives preference to decay channels(\ref{l2}, \ref
{l4}).

Remarkably, even with the small couplings (\ref{11}) and (\ref{12}) appeared
in HHLM model the LSP could decay inside a typical detector. In fact,
three--body decays of the LSP (\ref{l1}-\ref{l4}) drive at the widths \cite
{daw} 
\begin{equation}
\Gamma _{ikk}=\frac{\alpha c_{f}}{128\pi ^{2}}(\epsilon _{i}\frac{m_{f_{k}}}{%
Vc_{\beta }})^{2}\frac{M_{\chi _{1}^{0}}^{5}}{\widetilde{m}_{f}^{4}}
\label{17}
\end{equation}
where the corresponding effective LNV coupling constants were taken from
basic equations (\ref{u}) and (\ref{w}) for lepton ($f_{k}=e_{k},$ $c_{e}=1$%
) and quark ($f_{k}=d_{k},$ $c_{d}=3$ ) cases, respectively ($\widetilde{m}%
_{f}$ stands for masses of the intermediate sfermions involved, while $c_{f}$
is a color factor)). Assuming then that the LSP decays inside the detector ($%
c\gamma _{L}\tau (\chi _{1}^{0})\lesssim 1$ m, $\gamma _{L}$ is the Lorentz
boost factor) one obtains the lower bounds on the generic LNV parameters $%
\epsilon _{i}$%
\begin{equation}
\epsilon _{i}\gtrsim 10^{-6}\left( 2\gamma _{L}\frac{3}{c_{f}}\right)
^{1/2}\left( \frac{174GeV}{m_{f_{k}}}c_{\beta }\right) \left( \frac{%
\widetilde{m}_{f}}{200GeV}\right) ^{2}\left( \frac{100GeV}{M_{\chi _{1}^{0}}}%
\right) ^{5/2}  \label{lspd}
\end{equation}
Therefore, even for the possible smallest $\epsilon _{i}$ value, $\epsilon
_{i}$ $\sim 0.01$, as is likely to be conditioned by neutrino masses (see
next Subsection), one can expect to observe the LSP decays (including those
into light leptons and quarks) inside the detector, if sfermions are not
enormously heavy.

All the distinctive features of the LSP decays listed above taken together
constitute a main basis for a global testing of HHLM model. Some of
relations shown, such as (\ref{q}), \ref{qq}) and, especially, (\ref{z})
suggest the direct testing of the model, the rest allows to extract actual
values of the $\epsilon $-parameters to compare them with those extracted
from $\tau $ lepton decays (\ref{15}) or quite the reverse.

\subsection{Neutrino masses and oscillations}

The SUSY inspired lepton number violation opens in substance the shortest
way to neutrino masses and, indeed, many interesting attempts were made
towards this problem (see [7, 11, 25] and references therein).

In the HHLM model with the highly suppressed direct Higgs-lepton mixing
terms (\ref{sss}) one can neglect the tree-level neutrino masses and be
focused only on their radiative masses caused by the trilinear LNV couplings
(\ref{f}). Generally, they contribute to each entry of the neutrino Majorana
mass matrix through the diagrams with lepton--slepton and quark--squark
loops. It is apparent with the hierarchies in the basic LNV couplings(\ref{u}%
, \ref{w}) taken that the diagrams involving tau--stau and bottom--sbottom
loops are highly dominant. Therefore, to the obviously good approximation
the neutrino mass matrix comes finally to the remarkably transparent form 
\begin{equation}
M_{ij}^{\nu }\simeq \frac{3}{8\pi ^{2}}\frac{m_{b}^{2}}{V^{2}c_{\beta }^{2}}%
\frac{m_{b}^{2}}{\widetilde{m}_{b}^{2}}(A^{b}+\mu \tan \beta )\left( 
\begin{tabular}{lll}
$\epsilon _{1}^{2}(1+\Lambda )$ & $\epsilon _{1}\epsilon _{2}(1+\Lambda )$ & 
$\epsilon _{1}\epsilon _{3}$ \\ 
$\epsilon _{1}\epsilon _{2}(1+\Lambda )$ & $\epsilon _{2}^{2}(1+\Lambda )$ & 
$\epsilon _{2}\epsilon _{3}$ \\ 
$\epsilon _{1}\epsilon _{3}$ & $\epsilon _{2}\epsilon _{3}$ & $\epsilon
_{3}^{2}$%
\end{tabular}
\right)  \label{n3}
\end{equation}
where 
\begin{equation}
\Lambda =\frac{m_{\tau }^{4}}{3m_{b}^{4}}\frac{\widetilde{m}_{b}^{2}}{%
\widetilde{m}_{\tau }^{2}}\frac{A^{\tau }+\mu \tan \beta }{A^{b}+\mu \tan
\beta }\text{ ,}  \label{nn3}
\end{equation}
$\widetilde{m}_{\tau }$ and $\widetilde{m}_{b}$ stand for stau and sbottom
masses, while $A^{\tau ,b}$ are the corresponding trilinear soft terms in
the stau and sbottom left-right masses squared $m_{\tau ,b}(A^{\tau ,b}+$ $%
\mu \tan \beta )$, respectively.

Ignoring for the moment the relatively small contributions stemming from the
tau--stau loop ($\Lambda $ $\ll 1$) we come to one massive neutrino state
with mass 
\begin{equation}
m_{3}\approx 4.5\cdot 10^{-4}\frac{(A^{b}+\mu \tan \beta )}{\widetilde{m}%
_{b}^{2}c_{\beta }^{2}}(\epsilon _{1}^{2}+\epsilon _{2}^{2}+\epsilon
_{3}^{2})\text{$GeV$}^{2}  \label{n4}
\end{equation}
and two massless states. To account for the SuperKamiokande data \cite{2}
for $\Delta m_{atm}^{2}\approx 0.005$ $eV^{2}$ we require that 
\begin{equation}
\frac{(A^{b}+\mu \tan \beta )}{\widetilde{m}_{b}^{2}c_{\beta }^{2}}(\epsilon
_{1}^{2}+\epsilon _{2}^{2}+\epsilon _{3}^{2})\approx 1.6\cdot 10^{-7}\text{$%
GeV$}^{-1}.  \label{n5}
\end{equation}
Furthermore, according to the same data\cite{2} this massive state should be
about the maximal mixture of $\nu _{\mu }$ and $\nu _{\tau }$, while an
admixture of $\nu _{e}$ should be small. This requires the following
hierarchy\footnote{%
Actually, the CHOOZ data on $\nu _{e}$ disappearance \cite{CHOOZ} can be
accommodated with $\epsilon _{1}^{d}/\epsilon _{2,3}^{d}\lesssim 0.1$ (for a
coherent analysis of the neutrino mixings dictated by atmospheric and solar
neutrino data see \cite{BARB})} 
\begin{equation}
\epsilon _{1}\ll \epsilon _{2}\sim \epsilon _{3}.  \label{n6}
\end{equation}
The $\Lambda $ terms (tau--stau loop contributions) in $M_{ij}^{\nu }$(\ref
{n3}) result in some non-zero mass value $m_{2}\approx \Lambda m_{3}$ for
the next state, thus leaving finally only one neutrino state to be strictly
massless, $m_{1}=0$. Therefore, for the neutrino mass-squared differences
presently being of particular interest we have 
\begin{equation}
\frac{\Delta m_{sol}^{2}}{\Delta m_{atm}^{2}}=\frac{m_{2}^{2}-m_{1}^{2}}{%
m_{3}^{2}-m_{2}^{2}}\approx \frac{m_{2}^{2}}{m_{3}^{2}}\approx \Lambda ^{2}%
\text{ .}  \label{n7}
\end{equation}
Demanding that $\Lambda \approx 0.1$ (say, $\frac{\widetilde{m}_{b}}{%
\widetilde{m}_{\tau }}\approx 3$ and $A^{\tau }\approx A^{b}$ in Eq. (\ref
{n3})), one can account for the solar neutrino data in the small angle MSW
solution\cite{MSW} context. Thus, it looks like that both atmospheric and
solar neutrino data are well accommodated within the minimal LNV model with
the radiatively induced neutrino masses.

On the other hand, even a simple order of magnitude estimate shows (see Eq. (%
\ref{n5})) that to get the observed neutrino mass scale one must require a
decrease of the basic LNV parameters $\epsilon _{i}$ down to the order of $%
O(0.01)$ (unless the unnatural cancellation in neutrino mass scale (\ref{n4}%
) resulting in $A^{b(\tau )}+\mu \tan \beta =O(1)$ $MeV$) somehow occurs).
If so, then LNV interactions practically have no direct implications for
low-energy physics (such as those discussed in Subsections III.B, C) other
than the neutrino phenomenology, while they will still significantly alter
SUSY signal related with the LSP decays . Remarkably, even in this case,
since the parameters $\mu _{i}$ are turned out to be properly diminished
(see Eq. (\ref{sss})) the LNV-Yukawa alignment (\ref{u}, \ref{w}) continues
to work successfully, and with it all related predictions for the LSP decays
(Subsection III.D).

Finally, it is worthy of note that the simple structure of the neutrino mass
matrix (\ref{n3}) could somewhat be altered within the extended GUTs due to
the additional mixings of active neutrinos with those of sterile, generally
presented in higher GUT multiplets (for some attempts to accommodate
neutrino data within GUTs, see \cite{GUT_neut}).

\subsection{HHLM versus HLM}

Now, let us consider briefly an ordinary HLM (Higgs-Lepton Mixing) model,
another minimal framework for LNV that could be given solely by just the
generic bilinear terms $\mu _{i}L_{i}H_{u}$ in the superpotential $\Delta W$
(\ref{1}). Rotating them away one recovers the Yukawa-aligned trilinear LNV
couplings which, as those in the HHLM model (\ref{f}) , naturally overcome
all the currently available experimental constraints following from
low-energy physics (Subsection III.B).

However, the fundamental part of this scenario is related with SUSY
soft-breaking sector owing to which the condensation of sneutrinos and, as a
result, new sets of the physical (mixed) states both in gauge and Higgs
sector arise \cite{mix}. Because of this there appear some principal
differences with the HHLM model, which might manifest themselves at an
observational level.

The main point is that, while the effective trilinear LNV couplings are
generated in the HLM model,

\begin{equation}
\lambda _{ijk}=\xi _{i}Y_{jk}^{l}\text{ , \quad }\lambda _{ijk}^{\prime
}=\xi _{i}Y_{jk}^{d}\text{ \quad ( }\xi _{i}=\frac{<\widetilde{\nu }_{i}>}{%
Vc_{\beta }}-\frac{\mu _{i}}{\mu }\text{ )}  \label{hh}
\end{equation}
they, being properly weakened by the Yukawa couplings, appear too small even
for the dominant couplings\cite{chun}

\begin{equation}
\lambda _{i33}\approx 7\cdot 10^{-9}\frac{\eta }{c_{\beta }^{2}}\text{ ,
\quad }\lambda _{i33}^{\prime }\approx 2\cdot 10^{-8}\frac{\eta }{c_{\beta
}^{2}}\text{ \quad }  \label{hh1}
\end{equation}
(where $\eta =(M_{1}M_{2}/M_{Z}M_{\widetilde{\gamma }}-M_{Z}s_{2\beta }/\mu
)^{1/2}$ with $M_{1,2}$ standing for the $U(1)_{Y}$ and $SU(2)_{W}$
soft-breaking gaugino mass terms and $M_{\widetilde{\gamma }%
}=c_{W}^{2}M_{1}+s_{W}^{2}M_{2}$) for not marginally high values of $\tan
\beta $ . Hence, their contributions to the LNV processes relative to those
from the direct LNV admixtures in the physical neutralino and chargino
states are quite negligible. As a result, LNV processes in the HLM model,
being dominantly mediated by $W$ and $Z$ bozons, are in essence
family-independent (exclusive of the dependence on the generic mixing
parameters $\xi _{i}$ by themselves) in sharp contrast to HHLM model where
they essentially conditioned by quark and lepton mass hierarchy.

For example, the rare leptonic decays of $\tau $ considered in Subsection
III.C appear also in the HHL model. However, they practically do not depend
now on the final lepton masses. The proper relations between their
branchings follow from those of the HHLM\ model when taking in Eqs. (\ref{15}%
) all lepton masses equal (while $\epsilon _{i}/\epsilon _{j}\rightarrow \xi
_{i}/\xi _{j}$):

\begin{equation}
\Gamma (\tau \rightarrow 3e)=\Gamma (\tau \rightarrow e\mu \overline{\mu }%
)=\left( \frac{\xi _{1}}{\xi _{2}}\right) ^{2}\Gamma (\tau \rightarrow \mu e%
\overline{e})=\left( \frac{\xi _{1}}{\xi _{2}}\right) ^{2}\Gamma (\tau
\rightarrow 3\mu )  \label{c0}
\end{equation}

The same can be stated about the LSP decays as well (III.D). Their branching
relations are also largely quark and lepton mass-independent in the HHL
model. Therefore, again, instead of the hierarchical relations (\ref{lsprel}%
) one has the ''democratical'' ones

\begin{equation}
\frac{\Gamma (\chi _{1}^{0}\rightarrow e_{i}\overline{e}_{k}\nu _{k})}{%
\Gamma (\chi _{1}^{0}\rightarrow e_{j}\overline{e}_{m}\nu _{m})}\simeq \frac{%
\Gamma (\chi _{1}^{0}\rightarrow e_{i}\overline{d}_{p}u_{p})}{\Gamma (\chi
_{1}^{0}\rightarrow e_{j}\overline{d}_{q}u_{q})}\simeq \left( \frac{\xi _{i}%
}{\xi _{j}}\right) ^{2}  \label{c3}
\end{equation}
($i,$ $j$, $k,$ $m,$ $p,$ $q$ are any generation indices, no summing is
imposed) for decays (\ref{l1}, \ref{l3}) mediated now by $W$ bosons and
similar relations for decays (\ref{l2}, \ref{l4}) mediated by $Z$ bozon. The 
$Z$ bozon exchange gives one more peculiarity to the HLM model opening a way
to two new decay modes in the leptonic and semi-leptonic sector,
respectively. They are $\chi _{1}^{0}\rightarrow \nu _{i}\overline{\nu }%
_{k}\nu _{k}$ and $\chi _{1}^{0}\rightarrow \nu _{i}\overline{u}_{k}u_{k}$
coming solely from the neutrino ($\nu _{i}$) admixtures in the LSP.

As to the partial decay rates of the $\tau $ lepton and LSP in the HLM
model, they, according to the presently deduced constraints on the $\xi _{i}$
parameters coming from the existing bounds for lepton flavor-changing decays
of $Z$--bozon and neutrino masses(see, e.g., [31, 32]), are turned out to be
approximately in the same area as those in the HHLM model.

In closing one can summarize that, despite some generic similarity, the HHLM
and HLM models present two directly opposed observational possibilities,
each supplied with a quite clear signature manifesting itself in the flavor
hieararchy or flavor democracy of the final states produced.

\section{ Grand unification}

We argue in this Section that the HHLM\ model can naturally be embedded in
the grand unified framework, particularly in the $SU(7)$ model\cite{FFF}.
The preference given to the $SU(7)$ model over other grand unified schemes
is essentially determined by the missing VEV solution to the gauge hierarchy
problem which naturally appears in some $SU(N)$ GUTs starting from the $%
SU(7) $\cite{JGK}. This is shown to lead to a similar hierarchy of baryon vs
lepton number violation.

We discuss first briefly how one can come to the $SU(7)$ GUT. Towards this
end let us consider a general $SU(N)$ SUSY GUT with the simplest
anomaly--free set combination of the fundamental and 2-index antisymmetric
representations 
\begin{equation}
3\cdot \left[ (N-4)\overline{\Psi }^{A}+\Psi _{[AB]}\right]  \label{g1}
\end{equation}
($A,B=1,...,N$ are the $SU(N)$ indices) for the three quark--lepton
generations like as $3\cdot \left[ \overline{5}+10\right] $ in the prototype 
$SU(5)$ model. As to Higgs sector of the model there are an adjoint Higgs
multiplet $\Sigma _{B}^{A}$ responsible for the starting breaking of $SU(N)$
and conjugated pair of multiplets $H$ and $\overline{H}$ (being specified
later) where the ordinary electroweak doublets reside. Besides, there should
be $N-5$ scalar superfields $\varphi ^{r}$ and $\overline{\varphi }^{r}$ ($r$
$=1,...,N-5$ ) which break $SU(N)$ to $SU(5)$ by their own. It is also
expected that certain of the matter and / or Higgs superfields in the model
can carry charges of some protecting side symmetry like as an anomalous $%
U(1)_{A}$ , as in the case considered.

Now we suppose that all the generalized Yukawa couplings as the R-parity
conserving (ordinary up and down Yukawas), so R--parity violating ones
allowed by $SU(N)\otimes U(1)_{A}$ symmetry are given by the similar set of
the dimension-5 operators of the form ($i,j,k$ are the generation indices,
the SU(7) indices are omitted) 
\begin{equation}
{\cal O}_{ij}^{up}\propto \frac{1}{M_{P}}(\Psi _{i}\Psi _{j})(H\varphi )
\label{3}
\end{equation}
\begin{equation}
{\cal O}_{ij}^{down}\propto \frac{1}{M_{P}}(\overline{\Psi }_{i}\Psi _{j})(%
\overline{H}\varphi )  \label{5}
\end{equation}
\begin{equation}
{\cal O}_{ijk}^{rpv}\propto \frac{1}{M_{P}}(\overline{\Psi }_{i}\Psi _{j})(%
\overline{\Psi }_{k}\Sigma )  \label{5a}
\end{equation}
which can be viewed as an effective interactions generated through the
exchange of some heavy states with Plank scale order masses (they might be
treated as states inherited from the massive string modes)\footnote{%
The up-down hierarchy in the quark mass spectrum is assumed to be properly
given in this case by the VEV ratio of different extra scalars $\varphi $
involved in ${\cal O}_{ij}^{up}$ and ${\cal O}_{ij}^{down}$, respectively.}.
When being generated by an exchange of the same superheavy multiplet the
operators (\ref{5}) and (\ref{5a}) appear with the dimensionless coupling
constants to be properly aligned.

However, one must ensure first the suppression of the BNV interactions since
they are generated from the coupling (\ref{5a}) as well. The key idea here
is that the adjoint field $\Sigma $ involved in the RPV coupling (\ref{5a})
could develop the missing VEV pattern with zero color components: 
\begin{equation}
<\Sigma >=diag[0,0,0,a_{4},a_{5},...,a_{N}]V_{GUT}  \label{6}
\end{equation}
where $\sum_{k=4}^{N}a_{k}=0$. It is easy to verify that with such a basic
vacuum configuration in the model the baryon number violating part of RPV
interactions are projected out from the low energy effective superpotential.

As it was shown in the recent papers\cite{JGK} the missing VEV
configurations like (\ref{6}) naturally appear in some extended $SU(N)$ GUTs
from $SU(7)$ to solve the doublet-triplet splitting problem. In the minimal $%
SU(7)$ case\cite{FFF} which still remains an ordinary local symmetry of MSSM
at low energies the solution (\ref{6}) has a form

\begin{equation}
<\Sigma >=diag[0,0,0,1,1,-1,-1]V_{GUT}\text{ .}  \label{8}
\end{equation}
This breaks the $SU(7)$ symmetry to

\begin{equation}
SU(7)\rightarrow SU(3)_{C}\otimes SU(2)_{W}\otimes SU(2)_{E}\otimes
U(I)_{1}\otimes U(I)_{2}  \label{6''}
\end{equation}
while the extra symmetry $SU(2)_{E}\otimes U(I)_{1}\otimes U(I)_{2}$
breaking to the standard hypercharge $U(I)_{Y}$ appears due to the
additional fundamental scalars $\varphi ^{1,2}$ and $\overline{\varphi }%
^{1,2}$ (septets and anti-septets of $SU(7)$) mentioned above. They are
supposed to develop their VEVs along the ''extra'' directions 
\begin{equation}
\varphi _{A}^{1}=\delta _{A6}V_{1,}\text{ }\varphi _{A}^{2}=\delta _{A7}V_{2}
\label{8a}
\end{equation}
only through the proper Fayet-Iliopoulos $D-$term related with anomalous $%
U(1)_{A}$ symmetry \cite{witten} This is specially introduced in the $SU(7)$
model as a protecting symmetry which keeps extra symmetry-breaking scalars (%
\ref{8a}) untied from the basic adjoint scalar $\Sigma $ not to influence
the missing VEV solution appeared (\ref{8}).

One can readily check that a solution (\ref{8}) gives a minimum to a general
adjoint superpotential containing any even powers of $\Sigma $ (conditioned
by the reflection symmetry $\Sigma \rightarrow -\Sigma $ imposed): 
\begin{equation}
W_{A}=\frac{1}{2}m\Sigma ^{2}+\frac{\lambda _{1}}{4M_{P}}\Sigma ^{4}+\frac{%
\lambda _{2}}{4M_{P}}\Sigma ^{2}\Sigma ^{2}+...  \label{7}
\end{equation}
with $V_{GUT}\sim \frac{1}{\lambda }(mM_{P})^{1/2}$ which, for the properly
chosen adjoint mass $m$ and coupling constants $\lambda _{1,2,...}$, can
easily comes up to the string scale $M_{str}$. The superpotential $W_{A}$
can also be viewed as an ordinary renormalizable two-adjoint superpotential
with the second heavy adjoint scalar to be further integrated out.

Now, let us see how this missing VEV mechanism works to solve the gauge
hierarchy problem or, equivalently, the doublet-triplet splitting problem in
the $SU(7)$ SUSY GUT. There is, in fact, the only reflection-invariant
coupling of the basic adjoint $\Sigma $ with a pair of the ordinary
Higgs-boson containing supermultiplets $H$ and $\overline{H}$ 
\begin{equation}
W_{H}=f\overline{H}\Sigma H\text{ \hspace{0.5cm}}(\Sigma \rightarrow -\Sigma
,\overline{H}H\rightarrow -\overline{H}H)  \label{5b}
\end{equation}
having the zero VEVs, $H=$ $\overline{H}$ $=$ $0$, during the first stage of
the symmetry breaking. Thereupon $W_{H}$ turns to the mass term of $H$ and $%
\overline{H}$ depending on the missing VEV pattern (\ref{8}). This vacuum,
while giving generally heavy masses (of order of $M_{GUT}$) to them, leaves
their weak components strictly massless. To be certain we must specify the
multiplet structure of $H$ and $\overline{H}$ in the case of the
color-component missing VEV solution (\ref{8}) appeared in the $SU(7)$. One
can see that $H$ and $\overline{H}$ multiplet must be the 2-index
antisymmetric $21$-plets of $SU(7)$ which after starting symmetry breaking (%
\ref{6''}) contain just a pair of the massless weak doublets of MSSM (for
more detail see\cite{FFF}). Thus, there certainly is a natural
doublet-triplet splitting although we are coming to the strictly vanishing $%
\mu $-term at the moment. However, at the next stage when SUSY breaks,
radiative corrections shift the missing VEV to some nonzero value of order $%
M_{SUSY}$, thus inducing the ordinary $\mu $-term of MSSM\footnote{%
At this stage the effective bilinear LNV terms are also generated but they
are still suppressed relative to the ordinary $\mu $-term just as in the
case of MSSM disscussed in Section II (see Eq.(\ref{sss})).} , on the one
hand, and BNV couplings with the hierarchically small constants $\lambda
_{ijk}^{\prime \prime }=O(M_{SUSY}/M_{GUT})$, on the other.

Now, substituting the VEVs for scalars $\Sigma $ (\ref{8}) and $\varphi $ (%
\ref{8a}) in the basic operators (\ref{3}--\ref{5a}) , one obtains at low
energies the effective renormalizable Yukawa and LNV interactions, while the
BNV interactions are proved to be properly suppressed. Besides, if one
further introduces the superheavy intermediate $SU(7)$ septets $\Phi +%
\overline{\Phi }$, whose exchange generates the effective operators ${\cal O}%
_{ij}^{down}$(\ref{5}) and ${\cal O}_{ijk}^{rpv}$ (\ref{5a}) simultaneously,
the alignment between their dimensionless coupling of type (\ref{2}) follows
immediately. This can easily be read off the effective Yukawa and LNV
couplings(\ref{e}, \ref{f}) properly specified to the $SU(7)$ case ($%
\overline{H}_{d}\rightarrow \overline{H}$, $S\rightarrow \varphi $, $%
T\rightarrow \Sigma $).

The $SU(7)$ model is thoroughly considered in our forthcoming paper\cite{FFF}%
.

\section{Conclusions}

The recent neutrino data\cite{2} strongly suggest that neutrinos are
massive. While some other modifications of the SM could lead to neutrino
masses, SUSY extension of the SM with a generic lepton number violation
seems to be the most plausible and attractive framework.

In the present paper we have proposed some prototype model for a minimal LNV
which could appear in SUSY theories, if all the generalized Yukawa coupling,
both R-parity conserving and R-parity violating, had a common origin.While
our model is based on a generic Higgs-lepton mixing, we propose, in contrast
to the konwn picture, that this mixing appears not with a standard MSSM
Higgs doublet but with some superheavy weak doublets mediating the down
fermion (down quark and charged lepton) Yukawa couplings. As a result, all
significant physical manifestations of LNV are no other than those of the
effective trilinear couplings $LL\overline{E}$ and $LQ\overline{D}$ aligned,
by size and orientation in flavor space, with the down fermion effective
Yukawa couplings.

One of the immediate consequences of the HHLM model is a natural suppression
of the flavor-changing processes both in quark and lepton sectors due to the
additional flavor selection rules (\ref{bu}, \ref{ob}) appeared. According
to them a large reduction of a number of the possible LNV coupling
constants, from 36 of $\lambda _{ijk}$ and $\lambda _{ijk}^{\prime }$ to 3
of $\epsilon _{i}$ (\ref{ss}), takes place .

The model predicts a number of the potentially interesting signals (Section
III.B, C, D, E) which can be tested in future experiments. Experimental
study of the LSP decays is certainly of main interest for the model. These
decays, which even for the small LNV coupling constants could occur inside
the typical detectors, suggest the global testing of the HHLM model.

Simultaneously, we have shown that the present model leads to a
self-consistent picture of the radiatively induced neutrino masses and
mixings, thus successfully accommodating all the presently available
neutrino data, particularly, in the small angle MSW solution\cite{MSW}
context. However, the observed neutrino mass scale\cite{2} requires by
itself to suppress the magnitude of the LNV couplings demanding $\epsilon $%
-parameters down to the order of $10^{-2}$. If so, one can hardly expect any
experimentally interesting LNV signal in low--energy physics (Sections
III.B, C) beyond the neutrino phenomenology . Nevertheless, even in this
case the most significant HHLM predictions, which are related with the decay
modes of the LSP (Section III.D), remain in force.

Finally, we have presented the $SU(7)$ GUT framework for the minimal lepton
number violation. Remarkably enough, the SUSY inspired baryon number
violation is proved to be projected out from the low-energy superpotential
by the missing VEV vacuum configuration giving a solution to doublet-triplet
problem. So, a natural gauge hierarchy seems to lead to a similar hierarchy
of the baryon vs lepton number violation, at least, in the SUSY $SU(7)$ GUT.

\section*{Acknowledgments}

We should like to acknowledge the stimulating discussions with many our
colleagues, especially with Zurab Berezhiani, Grahame Blair, Daniel Denegri,
Gia Dvali, Colin Froggatt, Avto Kharchilava, Goran Senjanovic, Alexei
Smirnov and David Sutherland. This work was partially funded by a Joint
Project grant from the Royal Society. Financial support by INTAS Grants No.
RFBR 95-567 and 96-155 are also gratefully acknowledged.

\newpage

\section*{Tables}

\bigskip

\bigskip

\begin{tabular}{c}
\quad {\small TABLE I. Four-fermion operators resulting from LNV
interactions.}
\end{tabular}

\begin{tabular}{cccc}
\hline\hline
{\small Effective} & {\small Couplings} & {\small Particles} & {\small %
Example} \\ 
{\small operators} & {\small involved} & {\small exchanged} & {\small %
processes} \\ 
&  &  &  \\ 
$d_{j}\overline{d}_{k}d_{l}\overline{d}_{m}$ & $\lambda _{ijk}^{\prime
}\lambda _{iml}^{\prime \star }$ & $\widetilde{\nu }_{i}$ & $K^{0}-\overline{%
K}^{0}$, $B^{0}-\overline{B}^{0}$ \\ 
$u_{j}\overline{d}_{k}d_{l}\overline{u}_{m}$ & $V_{pj}^{\dagger
}V_{mq}\lambda _{ipk}^{\prime }\lambda _{iql}^{\prime \star }$ & $\widetilde{%
e}_{i}$ & $B\rightarrow K\pi $ \\ 
$u_{j}\overline{e}_{k}e_{l}\overline{u}_{m}$ & $V_{pj}^{\dagger
}V_{mq}\lambda _{kpi}^{\prime }\lambda _{lqi}^{\prime \star }$ & $\widetilde{%
d}_{i}^{c}$ & $\pi ^{0}\rightarrow \overline{\mu }e,$ $D^{+}\rightarrow \pi
^{+}\overline{\mu }e$ \\ 
$d_{j}\overline{e}_{k}e_{m}\overline{d}_{l}$ & $V_{ip}\lambda _{kpj}^{\prime
\star }\lambda _{mil}^{\prime }$, $\lambda _{ikm}^{\star }\lambda
_{ilj}^{\prime }$ & $\widetilde{u}_{i}$, $\widetilde{\nu }_{i}$ & $%
K^{0}\rightarrow e_{l}\overline{e}_{k}$, $K^{+}\rightarrow \pi ^{+}e_{l}%
\overline{e}_{k}$ \\ 
$d_{l}\nu _{j}\nu _{m}\overline{d}_{k}$ & $\lambda _{jil}^{\prime \star
}\lambda _{mik}^{\prime }$, $\lambda _{jki}^{\prime \star }\lambda
_{mli}^{\prime }$ & $\widetilde{d}_{i}$, $\widetilde{d}_{i}^{c}$ & $%
K^{0}\rightarrow \pi ^{0}\nu _{j}\overline{\nu }_{m},$ $B^{0}\rightarrow
K^{0}\nu _{j}\overline{\nu }_{m}$ \\ 
$u_{j}e_{k}\nu _{l}\overline{d}_{m}$ & $\lambda _{ijk}^{\prime }\lambda
_{iml}^{\prime },$ $\lambda _{ijk}\lambda _{iml}^{\prime }$ & $\widetilde{e}%
_{i}$, $\widetilde{d}_{i}$ & $\pi ^{-}\rightarrow \overline{\nu }_{l}\mu ,$ $%
B^{0}\rightarrow K^{+}e\overline{\nu }_{l}$ \\ 
$e_{l}\nu _{j}\nu _{m}\overline{e}_{k}$ & $\lambda _{ijk}\lambda
_{ilm}^{\star },$ $\lambda _{jki}\lambda _{mli}^{\star }$ & $\widetilde{e}%
_{i}$, $\widetilde{e}_{i}^{c}$ & $\mu \rightarrow e\nu _{j}\overline{\nu }%
_{m},$ $\tau \rightarrow \mu \nu _{j}\overline{\nu }_{m}$ \\ 
$e_{l}e_{j}e_{m}\overline{e}_{k}$ & $\lambda _{ijk}\lambda _{ilm}^{\star }$
& $\widetilde{\nu }_{i}$ & $\mu \rightarrow ee\overline{e},$ $\tau
\rightarrow ee\overline{\mu }$ \\ \hline\hline
\end{tabular}

\bigskip

\bigskip

\medskip

$
\begin{array}{l}
\text{{\small TABLE II. Three body decays of }}\mu \text{{\small and }}\tau 
\text{{\small , their expected branchings }} \\ 
\ \text{{\small and the current upper} {\small limits [18].}}
\end{array}
$

\begin{tabular}{ccc}
\hline\hline
{\small Decay process} & {\small Branching} & {\small Upper limit (CL=90\%)}
\\ 
&  &  \\ 
$\mu \rightarrow 3e$ & $7.2\cdot 10^{-18}\left( \frac{100GeV}{m_{\widetilde{%
\nu }_{2}}c_{\beta }}\right) ^{4}\left( \epsilon _{1}\epsilon _{2}\right)
^{2}$ & $<1.0\cdot 10^{-12}$ \\ 
$\tau \rightarrow 3e$ & $3.5\cdot 10^{-16}\left( \frac{100GeV}{m_{\widetilde{%
\nu }_{3}}c_{\beta }}\right) ^{4}\left( \epsilon _{1}\epsilon _{3}\right)
^{2}$ & $<2.9\cdot 10^{-6}$ \\ 
$\tau \rightarrow 3\mu $ & $1.5\cdot 10^{-11}\left( \frac{100GeV}{m_{%
\widetilde{\nu }_{3}}c_{\beta }}\right) ^{4}\left( \epsilon _{2}\epsilon
_{3}\right) ^{2}$ & $<1.9\cdot 10^{-6}$ \\ 
$\tau \rightarrow e\overline{e}\mu $ & $3.5\cdot 10^{-16}\left( \frac{100GeV%
}{m_{\widetilde{\nu }_{3}}c_{\beta }}\right) ^{4}\left( \epsilon
_{2}\epsilon _{3}\right) ^{2}$ & $<1.7\cdot 10^{-6}$ \\ 
$\tau \rightarrow \mu \overline{\mu }e$ & $1.5\cdot 10^{-11}\left( \frac{%
100GeV}{m_{\widetilde{\nu }_{3}}c_{\beta }}\right) ^{4}\left( \epsilon
_{1}\epsilon _{3}\right) ^{2}$ & $<1.8\cdot 10^{-6}$ \\ 
$\tau \rightarrow ee\overline{\mu }$ & {\small forbidden} & $<1.5\cdot
10^{-6}$ \\ 
$\tau \rightarrow \mu \mu \overline{e}$ & {\small forbidden} & $<1.5\cdot
10^{-6}$ \\ \hline\hline
\end{tabular}

\bigskip

\bigskip

\bigskip

\bigskip

\newpage

\end{document}